\documentstyle[12pt,aasms4]{article}
\tighten

\journalid{337}{}
\articleid{11}{14}

\def\spose#1{\hbox to 0pt{#1\hss}}
\def\lta{\mathrel{\spose{\lower 3pt\hbox{$\mathchar"218$}}
     \raise 2.0pt\hbox{$\mathchar"13C$}}}
\def\gta{\mathrel{\spose{\lower 3pt\hbox{$\mathchar"218$}}
     \raise 2.0pt\hbox{$\mathchar"13E$}}}
\def\n{\noindent}

\def\be{\begin{equation}}
\def\ee{\end{equation}}
\def\msun{M_{\odot}}

\begin{document}

\title {On a Theoretical Interpretation of the Period Gap in Binary 
Millisecond Pulsars} 
\author{Ronald E. Taam\altaffilmark{1}, Andrew R. King\altaffilmark{2},
Hans Ritter\altaffilmark{3}}

\n \altaffilmark{1}{Department of Physics \& Astronomy, Northwestern 
University, Evanston, IL 60208
\vskip -0.12 in
taam@apollo.astro.nwu.edu}

\n \altaffilmark{2}{Astronomy Group, University of Leicester, Leicester, 
LE1, 7RH, United Kingdom
\vskip -0.12 in
ark@star.le.ac.uk}

\n \altaffilmark{3}{Max Planck Institut f\"ur Astrophysik, Karl Schwarzschild
Str. 1, D-85740 Garching, Germany
\vskip -0.12 in
hsr@mpa-garching.mpg.de}

\begin{abstract}
We reexamine evolutionary channels for the formation of binary
millisecond pulsars in order to understand their observed orbital
period distribution. The available paths provide a natural division
into systems characterized by long orbital periods ($\gta 60$ days)
and short orbital periods ($\lta 30$ days). 
Systems with initial periods $\sim 1 - 2$ days, mainly driven by 
the loss of orbital angular momentum, ultimately produce low mass 
helium white dwarfs ($\lta 0.2 \msun$) with short orbital periods 
($\lta 1$ day). For longer initial periods ($\gta$ few days), early 
massive Case B evolution produces CO white dwarfs ($\gta 0.35 \msun$) 
with orbital periods of $\lta 20$ days. Common envelope evolutionary 
channels result in the formation of short period systems ($\lta 1$ day) from 
unstable low mass Case B evolution producing helium white dwarfs in 
the range $\sim 0.2 - 0.5 \msun$, and from unstable Case C evolution 
leading to CO white dwarfs more massive than $\sim 0.6 \msun$. On the  
other hand, the long orbital period group of binary millisecond pulsars
arise from stable low mass Case B evolution with initial orbital periods 
$\gta$ few days producing low mass helium white dwarfs and orbital periods 
$\gta 30$ days and from stable Case C evolution producing CO white dwarfs 
with masses $\gta 0.5 \msun$.   
The lack of observed systems between 23 and 56 days probably reflects the
fact that for comparable initial orbital periods ($\gta$ few days) 
low mass Case B and early massive Case B evolution lead to very discrepant 
final periods. We show in particular that the lower limit ($\sim 23$~d) 
cannot result from common--envelope evolution.

We discuss the importance of a phase of nonconservative evolution
where mass and angular momentum can be lost from the system through
the ejection of matter from accretion disks around the neutron stars
in these systems. This leads to a dependence of pulsar mass on
evolutionary history. In particular, most low--mass X--ray binaries
with orbital periods $\gta 2$~days are probably transient; the
super--Eddington accretion rates likely during outbursts mean that the
neutron stars in such systems gain relatively little mass.
\end{abstract}


\section{INTRODUCTION}

Ever since the discovery of the 1.55 ms radio pulsar PSR 1937+21 by
Backer et al.  (1982), the study of millisecond pulsars has yielded
fundamental insights into the properties of neutron stars and to their
evolution in binary star systems.  In order to determine the
statistical properties of the population, numerous radio surveys have
been undertaken recently covering both the northern hemisphere (e.g.,
Camilo 1999) and southern hemisphere (Manchester et al. 1996: Lyne et
al. 1998). The discovery that many pulsars of this class are members
of binary systems confirmed the hypothesis that the underlying neutron
star had accreted mass and been recycled to short spin period
(Radhakrishnan and Srinivasan 1982; Alpar et al. 1982). The recent
discovery of the 2.49 ms coherent pulse period in the X-ray pulsar SAX
J1808.4-3658 by Wijnands and van der Klis (1998) and the discovery of
its orbital period of 2.01 hours by Chakrabarty and Morgan (1998)
provided the first direct observational link between the two classes
of systems confirming that the progenitors of these systems are,
indeed, low mass X-ray binaries. For a review of the properties of the
millisecond binary pulsars see the paper by Phinney and Kulkarni
(1994).

A compilation of binary millisecond radio pulsars by Camilo (1995, 1999)
and more recent data reported in Manchester et al. (2000), Edwards 
(2000), and Tauris, van den Heuvel, \& Savonije (2000) reveals a 
population of 35 galactic binary pulsars with almost circular orbits 
and orbital periods ranging from about 2 hours to 175 days (see Table 1).  
It is found that the distribution of
systems with respect to orbital period is non uniform, and that there
is a significant absence of systems between about 23 and 56 days.
This is illustrated in Fig. 1 in the form of a cumulative distribution
of the orbital periods.  Since it is not likely that the absence of
systems in this period interval reflects a selection effect in their
discovery (Camilo 1995; Bhattacharya 1996), its origin must provide
insights to the evolutionary channels of these pulsar systems.
Recently, Tauris (1996) examined the possibility that the observed
"period gap" reflected the operation of a bifurcation mechanism in
their formation associated with distinct evolutionary histories.  In
particular, the long period systems were attributed to the evolution
of low mass systems along the red giant branch as shown by Joss and
Rappaport (1983), Paczynski (1983), and Savonije (1983) following
earlier work by Taam (1983) and Webbink, Rappaport, \& Savonije
(1983).  On the other hand, the short period systems were assumed to
be produced from short period (P $< 2$ days) low mass X-ray binary
systems which evolve to ever shorter orbital periods under the action
of angular momentum loss associated with magnetic braking (see Pylyser
\& Savonije 1988), and from long period (P $> 100$ days) X-ray
binaries which underwent a phase of common envelope evolution (van den
Heuvel and Taam 1984). In this picture, the pulsar companions are
primarily helium white dwarfs in the long period population and a
mixture of helium and carbon-oxygen white dwarfs in the short period
population.  The lower edge of the long period group (at $\sim$ 56
days) is attributable to the evolution of systems above the
bifurcation period dividing converging from diverging systems, see \S
2.1.1 below ($\sim 2$ days, Tauris 1996).  The upper period edge of
the short period group (at $\sim$ 20 days) is assumed to be the
maximum orbital period for systems emerging from the common envelope
phase.

In this paper, we reexamine these conjectures in light of the additional 
formation channel involving early massive Case B evolution recently identified 
by King \& Ritter (1999), and the current theory of common envelope evolution 
(Iben \& Livio 1993; Taam \& Sandquist 2000).  The description of the 
evolutionary paths as applied to the binary millisecond pulsars is presented 
in \S 2 and an interpretation of the orbital distribution and period gap in 
terms of the evolutionary processes is discussed in the final section.

After our paper was submitted (July 1999) two further papers (Tauris
\& Savonije, 1999; Tauris, van den Heuvel \& Savonije, 2000) relevant
to this subject appeared. We refer extensively to these papers
in what follows.

\section{EVOLUTIONARY MODELS}

In this section the contributions of the various formation channels to the 
binary millisecond pulsar stage are elucidated.  To place them into context
the individual paths are described in some detail.  Low mass Case B, early 
massive Case B, and common envelope evolution are presented in turn.

\subsection{Case B Evolution}

In this phase of evolution, mass transfer is driven by stellar expansion 
accompanying core contraction and the establishment of a hydrogen burning 
shell prior to the ignition of helium in the core.  The three distinct 
formation channels associated with this phase can be termed early low mass,
late low mass, and early massive evolutions.  Here, low mass refers to stars
($M < 2.25 \msun$) which develop a helium degenerate core and massive refers 
to stars developing a nondegenerate core.  

\subsubsection{Early Low Mass}

If mass transfer is initiated close to the main sequence, 
corresponding to initial orbital periods $\lta 1$ day, the binary evolution 
is affected by angular momentum loss processes associated with a magnetically 
coupled stellar wind (e.g., Verbunt \& Zwaan 1981).  The effectiveness of the 
tidal interaction between the components enforces synchronism, thus draining 
orbital angular momentum from the system. As a consequence, mass transfer is 
enhanced (producing a brighter low mass X-ray binary than would have been 
the case in the absence of such processes) and the system evolves to shorter 
orbital periods (Pylyser \& Savonije 1988).  Provided that the mass transfer 
is initiated for a sufficiently evolved star such that a degenerate core has 
formed the system can become detached producing a binary pulsar system 
(otherwise, the system continues its evolution as a low mass X-ray binary).  
For the masses relevant to the binary pulsar evolutions, the critical period 
(i.e., $P_{\rm bif}$ in the language of Pylyser \& Savonije 1988) is about 
0.4 - 0.8 days in the mass conservative case.  The actual value for the 
bifurcation period is dependent on the detailed assumptions concerning 
mass and angular momentum loss from the system.  Specifically, in the case 
of a strong consequential angular momentum loss the bifurcation period can be 
significantly longer than in the conservative case (e.g. Ergma et al. 1998).

\subsubsection{Late Low Mass}

In contrast to the above, binary stars with initial orbital periods greater 
than $P_{\rm bif}$ evolve into low mass X-ray binary systems with 
increasing orbital periods (Taam 1983; Webbink et al. 1983).  
Here, the orbital separation increases as a result of mass 
transfer from the less massive component to its neutron star companion. As 
the mass losing star evolves along the red giant branch the hydrogen rich 
envelope is depleted through a combination of nuclear evolution and mass loss, 
eventually leading to the formation of a white dwarf remnant in a wide detached
binary.  Since the radius of a giant is primarily a function of its core mass, 
the evolution can be described analytically (Ritter 1999).  Based on the 
core mass radius relation for Population I composition of Rappaport et al.
(1995), the relationship between the initial ($P_i$) and final ($P_f$) orbital 
periods, expressed in days, for mass transfer driven by nuclear evolution in 
the conservative mass transfer approximation is
$$
\log P_{\rm f}(d) = 1.31 + 0.70 \log P_{\rm i}(d) \eqno(1)
$$ 
where it has been assumed that the neutron star has an initial mass,
$M_{\rm 1,i} = 1.4 \msun$ and the low mass companion has an initial
mass $M_{\rm 2,i} = 1 \msun$.  On the other hand, if the mass transfer
process is non-conservative such that all the transferred mass is
ejected with the specific orbital angular momentum of the neutron
star, then the relationship with $M_1 =1.4 \msun$, $M_{\rm 2,i} = 1
\msun$ is
$$
\log P_{\rm f}(d) = 1.45 + 0.69 \log P_{\rm i}(d). \eqno(2)
$$
The correlation between the final orbital period 
and remnant white dwarf mass, $M_{WD}$, is given by 
$$
P_{\rm f} (d) \sim {8.5 \times 10^4 M_{\rm WD}^{5.5} \over (1 + M_{\rm
WD}^3 + 1.75 M_{\rm WD}^4)^{1.5}} \eqno(3)
$$ 
(see King \& Ritter 1999) where the white dwarf mass is expressed in
$\msun$. For Population II composition (which may be appropriate for
millisecond pulsars with spindown ages exceeding several Gyr), the 
corresponding eqn. 6 of Rappaport et al. (1995) leads to periods which
are smaller than the estimate (3) by about a factor 2.

We note that Tauris \& Savonije (1999) have recently recalculated
low mass Case B evolution using newer input physics. In particular
this produces binary millisecond pulsars with $P_{\rm orb} \simeq 1$~d
and $M_{\rm WD} \simeq 0.2\msun$, which were missing from the results
of Rappaport et al. (1995).
 
\subsubsection{Early Massive}

This type of evolution (see Kippenhahn \& Weigert 1967) describes a
binary in which the mass losing component has a radiative envelope and
is sufficiently massive that helium is ignited under nondegenerate
conditions (M $\gta 2.25 \msun$).  As applied to binary pulsars (see
King \& Ritter 1999) any mass transferred to the neutron star at rates
above the Eddington limit is assumed to be ejected from the inner
regions of an accretion disk in the form of a wind.  This implies that
the mass of the neutron star is little changed during this thermal
timescale phase.  The mass transfer is interrupted either by the
complete loss of the envelope or by the contraction of the
nondegenerate component within its Roche lobe when helium is ignited
at its center.  In these evolutions, the ultimate fate of the binary
is dictated by the initial mass ratio of the donor to the gainer.  For
initial mass ratios $q = M_{\rm donor}/M_{\rm NS}$ slightly smaller
than a critical mass ratio (which depends on the donor star's initial
mass and on the initial orbital separation), the evolving star
detaches from its Roche lobe with a thick hydrogen rich envelope.
Further evolution of the star to its double shell burning phase leads
to the resumption of mass transfer and to the expansion of the orbit.
The depletion of the envelope on the asymptotic giant branch leads to
the contraction of the core and to the formation of a CO white dwarf
with a neutron star companion in a long period orbit ($\lta 50-70$~d,
cf figs 2 and 4 of Tauris et al., 2000).
For the critical mass ratio the donor detaches from its Roche
lobe with a thin hydrogen envelope.  Near the end of mass transfer
such an evolution leads to a system resembling Cyg X-2, with an
orbital period of $\sim 10$ days and to the eventual formation of a
binary pulsar system consisting of a CO white dwarf and a neutron star
similar to J0621+1002 (King \& Ritter 1999).  Finally, for systems
characterized by initial mass ratios much greater than the critical
value, mass transfer removes the entire hydrogen rich envelope and the
remnant helium star directly evolves to form a CO white dwarf.  For
this type of evolution, larger initial mass ratios result in shorter
orbital periods for the binary pulsar system and to more massive CO
white dwarf companions (King \& Ritter 1999).  An approximate relation
for the minimum final period expressed in days for such systems is
given by
$$
\log P_{\rm f} \approx 2.56 - 3.1 \frac{M_{\rm WD}}{\msun}
\eqno(4)
$$
yielding, for example, a maximum of $\sim 30$ days for the minimum 
mass of a CO white dwarf ($0.35 \msun$) produced in this evolution.

\subsection{Common Envelope Evolution}

Systems evolving through the common envelope phase can also contribute
to the binary millisecond pulsar population (see Taam \& Sandquist
2000).  In such systems the mass transfer is unstable, leading to the
formation of a differentially rotating common envelope.  This
evolution is particularly relevant to binary systems characterized by
a large mass ratio and a giant star component.  For sufficiently large
mass ratios, the binary is subject to a tidal instability (Darwin
1879; Kopal 1972; Counselman 1973; Hut 1980) occuring when the orbital
angular momentum of the binary is less than 3 times the spin angular
momentum of the giant star. For orbital separations less than a
critical value, where the total angular momentum is a minimum, the two
stars must spiral together.  Progenitor systems can also evolve into a
common envelope provided that the envelope of the nondegenerate
component is convective.  In this case, the envelope can expand upon
mass loss for evolutionary stages where the fractional mass in the
convective envelope exceeds 0.5 (Hjellming 1989).  For mass ratios
$\gta 1.2 - 1.5$ a giant cannot be constrained within its Roche lobe
and a runaway mass transfer episode is expected.
 
Successful ejection of the envelope without requiring the two stars to
merge can be achieved given that sufficient energy is released from
the orbital motion and that the mass losing star is sufficiently
evolved (see Taam \& Sandquist 2000).  The outcome of such an
evolution would resemble a short--period binary pulsar like PSR
0655+64 (van den Heuvel \& Taam 1984).  For an interaction of a
neutron star with a companion on the first ascent red giant branch a
helium white dwarf companion of mass $\gta 0.2 - 0.25 \msun$ is
expected (Sandquist, Taam, \& Burkert 2000; see also Sarna, Marks, \&
Smith 1996). On the other hand, an interaction on the asymptotic giant
branch leads to a CO white dwarf with mass $\gta 0.55 \msun$. We note
that the evolution into a common envelope phase is less likely for a
star characterized by a radiative envelope (since the star contracts
upon mass loss) and for mass ratios not significantly larger than
unity (where the mass ratio can be reversed, leading to orbital
expansion).

The upper limit to the orbital period for the short-period population
of binary millisecond pulsars has been identified with the outcome of
common envelope evolution (Tauris 1996).  Although evolution via a
common envelope may be expected to populate the region at short
orbital periods ($\lta$ a few days), there is insufficient orbital
energy to produce a system with a period as long as 20 days for the
intermediate mass progenitors characteristic of the white dwarf
component in the system.  To circumvent this difficulty, Tauris (1996)
had to increase the efficiency of mass ejection in the common envelope
phase significantly above unity (i.e. to 4).  At the time the cutoff
was 12 days: the current value of 23 days (J1618-3919) exacerbates the
problem if we assume that this system also evolved through a CE phase.
Tauris (1996) invoked unspecified sources of energy to accommodate
this choice; however, the ionization energy and nuclear energy sources
that have been invoked in the past are ineffective (see Sandquist et
al. 2000).  Specifically, the ionization energy cannot be tapped
directly (see Harpaz 1998) and is unimportant for ejecting matter in
the deep gravitational potential of the binary where the major phase
of mass ejection takes place.  Nuclear energy is tapped on a timescale
too long compared to the mass ejection timescale to be important.

One might try to get round this argument by assuming that systems such
as J1618-3919 have low--mass helium white dwarf companions. 
However
this is ruled out because the fairly large values of $P_{\rm spin}$
(32.82~ms, 71.09~ms) for J1810-2005 and J1904+04 ($P_{\rm orb} =
15.0$~d and 15.75~d respectively), as well as the large value of
$\dot P_{\rm spin}$ for J1810-2005, strongly indicate that these
systems passed through a phase of very high mass transfer and that 
the neutron star cannot have accreted much matter. Since CE
evolution is probably ruled out by the argument of the last paragraph,
we conclude that the short $P_{\rm orb}$ population below the gap, but 
longer than a few days must be the result of early massive Case B evolution.

\section{DISCUSSION}

Given the possible evolutionary channels for the formation of binary
millisecond pulsars presented in the previous section, we consider
their implications for the period distribution of these systems.  As
described in \S 2, it is unlikely that the common envelope path is
responsible for the period cutoff of the short period systems and therefore
that the efficiency for mass ejection greater than unity is required. 
We suggest that the cutoff near 23 days is a natural consequence of the
early massive Case B evolution near the critical mass ratio (see King
\& Ritter 1999). The lower period cutoff for the long period systems
is likely to result from low mass Case B evolution, but it is not
related to the bifurcation mechanism favored by Tauris (1996) since a
bifurcation period at 2 days requires angular momentum loss rates
significantly greater than commonly believed.  On the contrary, we
suggest that this period distribution is a natural consequence of the
small phase space available to those systems which evolve into the 20
- 60 day period interval.  That is, although the formation of binary
millisecond pulsars with orbital periods as short as 20 days is
possible in this evolutionary channel, it is not likely, since the
initial core mass of the progenitor must be very low, i.e. $\sim 0.17
\msun$.  The initial orbital period corresponding to this evolution is
$\sim$ 1 day. As pointed out by Tauris (1996) such systems evolve to
shorter orbital periods because of angular momentum losses associated
with a magnetized stellar wind (Pylyser and Savonije 1988).  For those
systems which do not evolve to shorter periods, the final period is
extremely sensitive to the initial period at which mass transfer is
initiated.  For example, small initial core masses ($\sim 0.2 \msun$)
yield orbital periods $\gta 60$ days for initial orbital periods of
about 3 days.  We suggest that the small range of core masses means
that only a few systems can finish their evolution with a period in
the range of 20 -- 60 days.


Implicit in the evolutionary scenarios we discuss, and especially
important for understanding the short period population ($\sim 1 - 20$
days), is the assumption that the neutron star accretes only a small
fraction of the transferred mass because accretion on to the neutron
star is likely to occur at super--Eddington rates. This occurs either
because the mass transfer timescale is $\lta 10^8$ years or because
matter is accreted at high rates in a transient manner. This latter
circumstance probably results from a thermal instability in the
accretion disk (King, Kolb, \& Burderi 1996; King, Kolb, \&
Szuszkiewicz 1997; Li, van den Heuvel, \& Wang 1998).  The orbital
separations of long--period low--mass X--ray binaries (LMXBs) imply
very large accretion disks, whose outer regions cannot be maintained
in the hot (ionized) state, even by self--irradiation via
X--rays. Thus all LMXBs with $P_{\rm orb} \gta 2$~d are probably
transient (King, Kolb \& Burderi, 1996). In this case, accretion on to
the neutron star occurs only during outbursts.  A short outburst duty
cycle, as inferred from observation, implies super--Eddington central
accretion rates. Only a small fraction is accreted, while the
remainder is ejected as a wind. Calculations neglecting this mass loss
predict final neutron star masses systematically larger than the upper
limits on the observed masses, often by several times $0.1\msun$, cf
Fig. 4b of Tauris \& Savonije (1999).

Ejection is unlikely at short periods because most neutron star
systems are persistent, although transients do exist, e.g., SAX
J1808.4-3658 (Chakrabarty \& Morgan 1998) at 2 hours and EXO 0748-676
(Parmar et al. 1986) at 3.8 hours (cf King et al. 1996; King \& Kolb
1997).  In this case, the neutron star gains mass and is likely to be
spun up to millisecond periods.

Given the circumstances under which mass can be transferred at high
rates, one can expect that the mass of the pulsar is dependent on the
particular evolutionary path leading to its formation.  As the mass
transfer rates driven by nuclear evolution of highly luminous giant
stars exceed the Eddington rate, the neutron stars in the long period
population are not likely to have accreted much matter.  That is, the
neutron star components in those systems which have been formed via
the low mass Case B and intermediate early massive Case B evolutionary
channels are likely to have accreted little mass, and should differ
little from their formation masses.  Likewise, the neutron stars that
evolve through a common envelope phase are also not likely to have
accreted much matter. Although it has been hypothesized that they can
accrete sufficient matter in the neutrino dominated regime (Zeldovich,
Ivanova \& Nadyozhin 1972) to form black holes (Chevalier 1993; Brown
1995), recent studies (Chevalier 1996) indicate that neutron stars are
expected to survive when account is taken of rotational effects in
highly evolved giants.  In this context, the more evolved state of the
giant companion makes it less likely that the neutrino accretion
dominated regime is reached.  The lower rates of accretion
characteristic of the lower density envelopes of more evolved stars,
in combination with the short evolutionary phase of the common
envelope ($\lta 10 - 100$ years) suggest that accretion is not
important.  On the other hand, pulsars in systems with short orbital
periods ($\lta 1$ day) should be more massive than the other pulsars
in their class since the mass transfer rates are less than the
Eddington value.  The observational measurement and improved estimates
of pulsar masses in such systems following up on the work by Thorsett
\& Chakrabarty (1999) is highly desirable for testing such an
evolutionary dependence.

\acknowledgements

\n This research was initiated at the Institute of Theoretical Physics
and was supported in part by the National Science Foundation under
Grant No. PHY94-07194, and at Northwestern University by NSF grant
AST-9727875.  ARK also acknowledges support as a PPARC Senior Fellow.

\vspace*{-.5in}
\begin{deluxetable}{rrrrrrr}
\tablecolumns{7}
\tablewidth{0pc}
\tablecaption{Galactic ms-pulsars and pulsars in (circular) binaries}
\tablehead{
\colhead{PSR} & \colhead{$P_{\rm spin}(ms)$} & 
\multicolumn{2}{c}{$\dot P$} 
& \colhead{$P_{\rm orb}(d)$} & 
\colhead{$e$} & \colhead{($\frac{M_2}{\msun}$) \tablenotemark{a}}}
\startdata
J2051-0827  &   4.50864        &   1.27&$10^{-20}$&    0.09911      &$<$0.00008   &       0.0316       \nl
J0751+1807  &   3.47877        &   8.0 &$10^{-21}$&    0.26314      &$<$0.0001    &       0.153        \nl
B1957+20    &   1.60740        &   1.69&$10^{-20}$&    0.38197      &$<$0.00004   &       0.0254       \nl
J1756--5322 &   8.87           &\multicolumn{2}{c}{ }& 0.453        &             &       0.682        \nl
J1012+5307  &   5.25575        &   1.46&$10^{-20}$&    0.60467      &$<$0.00002   &       0.128        \nl
B0655+64    & 195.67094        &   6.86&$10^{-19}$&    1.02867      &   0.00000075&       0.813        \nl
J0613--0200 &   3.06184        &   9.57&$10^{-21}$&    1.19851      &   0.000007  &       0.153        \nl
J1435--60   &   9.35           &\multicolumn{2}{c}{ }& 1.355        &   0.00      &       1.11         \nl
J0034--0534 &   1.87782        &   0.7 &$10^{-20}$&    1.58928      &$<$0.00002   &       0.169        \nl
B1831--00   & 520.95431        &   1.43&$10^{-18}$&    1.81110      &   0.0001    &       0.075        \nl
J1232--6501 &  88.28           &   1.0 &$10^{-18}$&    1.863        &   0.00      &       0.175        \nl
J0218+4232  &   2.32309        &   8.00&$10^{-20}$&    2.02885      &$<$0.00002   &       0.200        \nl
J2317+1439  &   3.44525        &   4.42&$10^{-21}$&    2.45933      &   0.0000005 &       0.206        \nl
J1911--1114 &   3.62575        &   1.34&$10^{-20}$&    2.71656      &$<$0.000013  &       0.143        \nl
J1157--5112 &  43.6            &$<$0.9 &$10^{-18}$&    3.507        &             &       1.48         \nl
J1045--4509 &   7.47422        &   1.74&$10^{-20}$&    4.08353      &   0.000024  &       0.190        \nl
J0437--4715 &   5.75745        &   5.73&$10^{-20}$&    5.74104      &   0.000019  &       0.168        \nl
J1603--7202 &  14.84195        &   1.4 &$10^{-20}$&    6.30863      &$<$0.00002   &       0.346        \nl
J2129--5721 &   3.72635        &   2.0 &$10^{-20}$&    6.62549      &$<$0.000017  &       0.158        \nl
J2145--0750 &  16.05242        &   2.9 &$10^{-20}$&    6.83890      &   0.000018  &       0.514        \nl
J1022+1001  &  16.45293        &   4.2 &$10^{-20}$&    7.80513      &   0.000098  &       0.870        \nl
J0621+1002  &  28.85386        &$<$0.8 &$10^{-19}$&    8.31868      &   0.002458  &       0.539        \nl
J1804--2717 &   9.34303        &   4.2 &$10^{-20}$&   11.12871      &   0.000035  &       0.240        \nl
B1855+09    &   5.36210        &   1.78&$10^{-20}$&   12.32717      &   0.000027  &       0.290        \nl
J1453--58   &  45.25           &\multicolumn{2}{c}{ }&12.422        &   0.00      &       1.07         \nl
J1810--2005 &  32.82           &   1.3 &$10^{-19}$&   15.012        &   0.00      &       0.341        \nl
J1904+04    &  71.09           &\multicolumn{2}{c}{ }&15.750        &   0.04      &       0.270        \nl
J1709+23    &   4.631          &\multicolumn{2}{c}{ }&22.7          &             &                    \nl
J1618-3919  &                  &\multicolumn{2}{c}{ }&23.3          &             &                    \nl
J2033+17    &   5.94896        &\multicolumn{2}{c}{ }&56.2          &$<$0.05      &       0.223        \nl
J1713+0747  &   4.57014        &   8.52&$10^{-21}$&   67.82513      &   0.000075  &       0.332        \nl
J1455--3330 &   7.98720        &$<$0.6 &$10^{-21}$&   76.17458      &   0.000167  &       0.304        \nl
J2019+2425  &   3.93452        &   7.02&$10^{-21}$&   76.51163      &   0.000111  &       0.373        \nl
J2229+2643  &   2.97782        &   1.46&$10^{-21}$&   93.01590      &   0.000256  &       0.146        \nl
B1953+29    &   6.13317        &   2.97&$10^{-20}$&  117.34910      &   0.000330  &       0.231        \nl
J1643--1224 &   4.62164        &   1.84&$10^{-20}$&  147.01740      &   0.000506  &       0.142        \nl
J1640+2224  &   3.16332        &   2.83&$10^{-21}$&  175.46066      &   0.000797  &       0.295        \nl
J1803--2712 & 334.41542        &   1.73&$10^{-17}$&  406.781        &   0.000507  &       0.172        \nl
B0820+02    & 864.87275        &   1.04&$10^{-16}$& 1232.47         &   0.011868  &       0.231        \nl
\enddata
\tablenotetext{a}{$\frac{M_2}{\msun}$ is calculated assuming a pulsar 
mass of $1.4 \msun$ and a system inclination angle of 60 degrees.}
\end{deluxetable}

\vfil\eject

\begin{figure}[t] 
\plotone{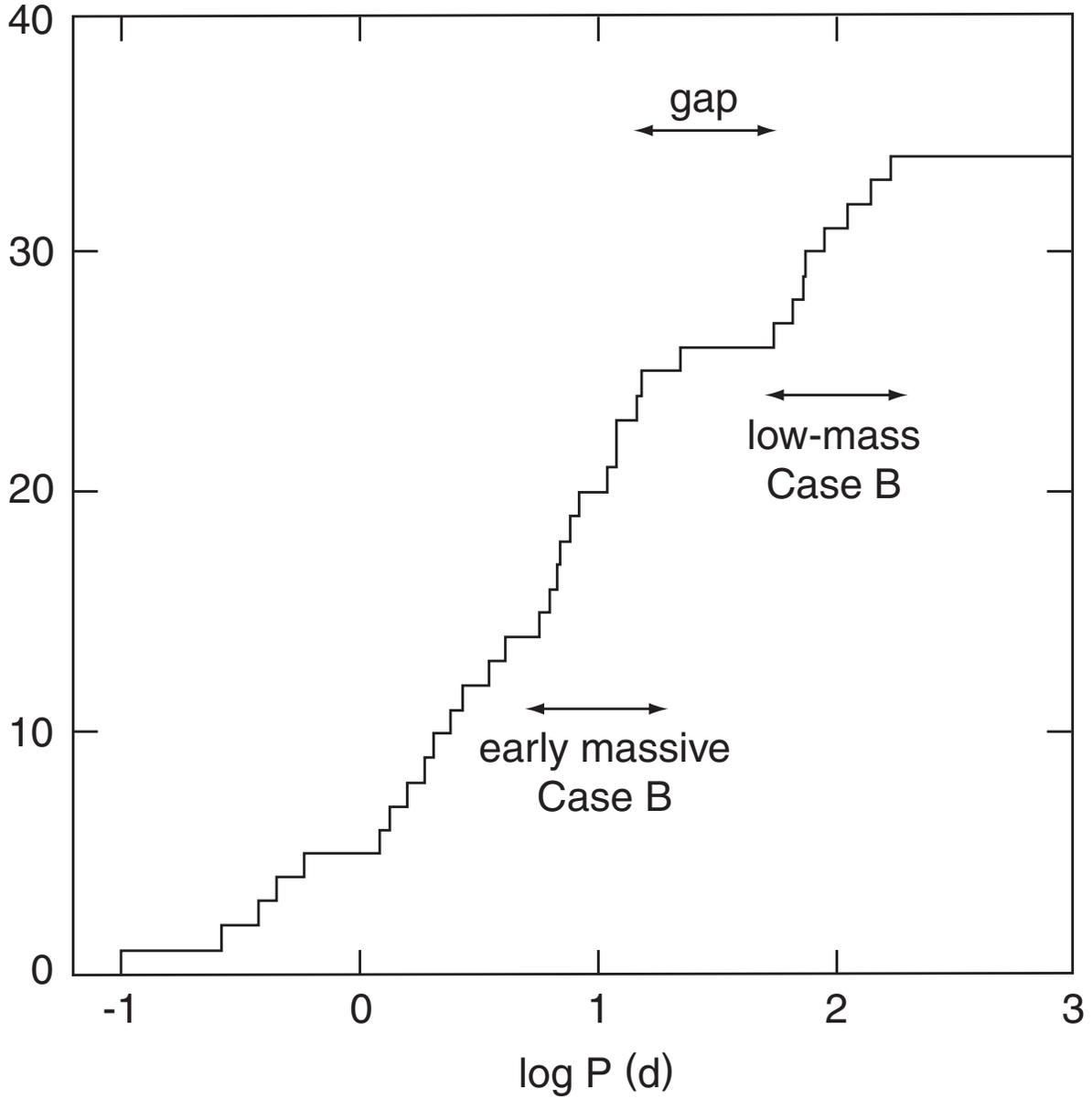}
\caption[]{The cumulative distribution of orbital periods of known binary 
galactic millisecond pulsars with circular orbits.  The gap region in which 
there exists one object between 12 and 56 days is indicated.  Also shown are 
the major evolutionary channels contributing to systems just below and above
the gap (see text). }
\end{figure}

\vfil\eject

\end{document}